\documentclass[twocolumn,showpacs,prl,amsmath,amssymb]{revtex4}
\usepackage{graphicx}
\usepackage{dcolumn}
\usepackage{bm}

\begin{document}
  
\title{Boosting electronic transport in carbon nanotubes by isotopic disorder}
\author{Niels Vandecasteele, Michele Lazzeri and Francesco Mauri}
\affiliation{IMPMC, Universit\'es Paris 6 et 7, CNRS, IPGP, 140 rue de Lourmel, 75015 Paris, France}
\begin{abstract}
The current/voltage curve of metallic carbon nanotubes (CNTs) displays at high bias a sudden increase of the resistivity due to the scattering of electrons with phonons having an anomalously-high population (hot phonons).  Here, we show that it is possible to improve the electrical performances of metallic CNTs by $^{13}$C isotope enrichment.  In fact, isotopic disorder creates additional channels for the hot-phonon deexcitation, reduces their population and, thus, the nanotube high-bias differential-resistance.  This is an extraordinary case where disorder improves the electronic transport.

\end{abstract}

\maketitle

\paragraph*{Introduction.---}
Metallic single-wall carbon nanotubes (CNTs) are quasi one-dimensional
wires which can carry the highest current density
(${10^9}$~A/cm$^{-2}$)~\cite{yao00} of any material. This makes them
the best candidates as interconnects in future high-power
electronic-devices.
The measured current/voltage (IV) curve of metallic single-wall
CNTs with ohmic contacts~\cite{yao00,javey04,park04} displays at high
bias (voltages higher than $\sim$0.2 Volts) a sudden increase of the
resistivity, due to the scattering of conducting electrons with
optical atomic vibrations (phonons)~\cite{yao00}.
The largest part of this resistivity is
due to the presence of an anomalously-high optical-phonon occupation
(hot phonons)~\cite{lazzeri05,lazzeri06a}.  Indeed, the rate at which
optical phonons are excited (by the scattering with conducting
electrons) is faster than the rate at which they are deexcited.
The large hot-phonon population determines an important
part of the electrical resistivity, limiting the performances of CNTs.  
So far, the most direct experimental observations of the phenomenon 
are given by Refs.~\cite{oron-carl08,bushmaker07},
where an important increase of the
optical-phonon population during electron transport is measured by
Raman spectroscopy.
Also, in Ref.~\cite{pop05} a negative differential resistance
in tubes suspended between the two electrodes is interpreted as due
to hot phonons.

In this paper, we show that it is possible, in practice, to improve
the electrical performances of metallic nanotubes by $^{13}$C isotope
enrichment.
Isotopic substitution changes the nuclear masses but not the
nuclear charges and thus introduces elastic scattering mechanisms
for phonons but not for electrons
(i.e. it does not degrade the electronic transport
properties)~\cite{note-1}.
In fact, the introduction of isotopic disorder creates
additional deexcitation channel for the hot phonons.
This results in a reduction of the hot-phonon population
and in an important decrease of the nanotube high-bias
differential-resistance. The phenomenon is  quantified by means of a
coupled Boltzmann transport equation (BTE) for both phononic and electronic
populations, in which all the relevant scattering parameters are
obtained from ab-initio calculations based on density functional
theory (DFT).

\begin{figure}[h!]
\centerline{\includegraphics[width=80mm]{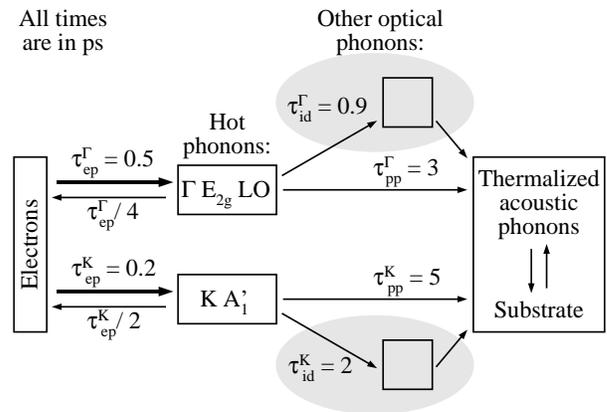}}
\caption{Schematic view of the hot phonon generation (see the text).
In a defect-free tube, the hot phonons are generated and thermalized
with characterisitc times $\tau_{ep}$ and $\tau_{pp}$, respectively.
Isotopic disorder introduces additional
thermalization channels (shaded area)
with characteristic times $\tau_{id}$.}
\label{fig1}
\end{figure}

\paragraph*{Interplay between electron-phonon scattering and thermalization.---}
The scattering processes involved in the degradation of the
electronic transport are represented in Fig.~\ref{fig1}.
In metallic CNTs,
conducting electrons can scatter mainly with two optical phonons,
corresponging in graphene to the longitudinal E$_{2g}$
phonon near ${\bm \Gamma}$ (${\bm \Gamma}$E$_{2g}$LO) and to the
A$'_1$ near {\bf K} ({\bf K}A$'_1$)~\cite{note00}.
Their momentum is localized in a
small region of the Brillouin zone (BZ)~\cite{lazzeri06a,note00}
and their energy is $\sim$0.2~eV. For
voltages higher than 0.2 Volts, conducting electrons have
sufficient energy to excite the two phonons. As a consequence, the
electron scattering and, thus, the resistance increase.  For tubes
with a diameter of 2~nm (typically found in
experiments) the ${\bm \Gamma}$E$_{2g}$LO and {\bf K}A$'_1$ phonons are
excited with a characteristic electron-phonon scattering time
$\tau_{ep}^{\bm \Gamma}\sim0.5$ and $\tau_{ep}^{\bf K}\sim0.2$~ps,
respectively~\cite{note01}. In a defect-free tube,
the two phonons are deexcited (or thermalized) with 
characteristic times $\tau_{pp}^{\bm \Gamma}\sim3$ and $\tau_{pp}^{\bf
K}\sim5$~ps which are
due to anharmonic phonon-phonon scattering~\cite{note01}.
In fact, anharmonicity couples the two phonons with acoustic phonons
which can be considered thermalized in the typical experimental
conditions (when the tubes are lying on a substrate)~\cite{note_ref01}.
In practice, the excitation of the two optical phonons is faster than
their thermalization ($\tau_{ep} \ll \tau_{pp}$). Because of this,
the two optical phonons can not be thermalized with the rest of the tube
and their occupation is expected to increase during high-bias
electronic transport (hot phonons).  This, in turn, will result in a
larger electron-phonon scattering and a larger electronic resistance.

\begin{figure}[h!]
\centerline{\includegraphics[width=75mm]{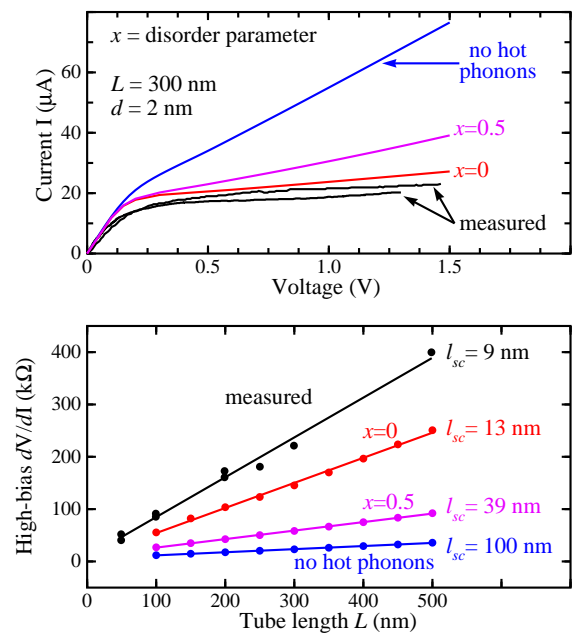}}
\caption{(Color online) Upper panel: IV curve of a (14,14) nanotube
with length $L$ and diameter $d$. 
Black lines are measurements from Refs.~\cite{javey04,park04}.
Red and violet lines are calculations for an isotopically pure ($x=0$)
and enriched tube ($x=0.5$).
The blue line is obtained by imposing the phonon populations to be zero.
Lower panel: high-bias differential resistance for (14,14) tubes
with different lengths (same color code as in the upper panel).
Lines are linear fit to the points and
$l_{sc}$ is the resulting scattering length (Eq.~\ref{eq4}).}
\label{fig2}
\end{figure}

A quantitative description of the process can be obtained by means
of a coupled BTE for both phononic and
electronic populations, as in Ref.~\cite{lazzeri06a}
(see note ~\cite{note02}).
The resulting IV curve reproduces very well experiments
(compare the red curve with measurements in  Fig.~\ref{fig2},
upper panel).  The red curve of Fig.~\ref{fig2} is
obtained by letting the phonon population evolve according to BTE and
is associated to an important increase of the optical phonons
population.  To show the relevance of this, in
Fig.~\ref{fig2} we also show the IV curve obtained by imposing the
phonon population to be zero (blue curve). 
As already remarked in Ref.~\cite{lazzeri06a}, in this case the
high-bias current is much higher, that is {\it an important part of
the measured high-bias resistance is due to the presence of hot
phonons}.
We stress that in the calculations of Fig.~\ref{fig2} all the
relevant parameters are obtained from DFT and no parameters are present
which can be tuned in order
to recover agreement with measurements~\cite{note02}.

The bottleneck for hot-phonon generation is
the large phonon thermalization-time.  Thus, the set up of
a mechanism to decrease it will diminish the
hot-phonon population improving the electron conduction.
This goal could be reached by changing the isotopic
composition of the nanotubes, since isotopic disorder provides
a new phonon-phonon scattering mechanism.
On the other hand, the possibility
of synthesizing isotopically enriched nanotubes has been actually
demonstrated ~\cite{BN,Kuzmany1,Kuzmany2}.

In a defect-free crystal, a phonon is
characterized by a momentum {\bf q} and a branch index $\nu$.  When
isotopic disorder can be considered as a perturbation, the dynamic of
the system can be described by the evolution of the
unperturbed {\bf q}$\nu$ states.  Isotopic
disorder results in the creation of additional scattering channels
between the phonons, characterized by a time $\tau_{id}$
(Fig.~\ref{fig1}).  In particular, an initial hot-phonon
${\mathbf{q}\nu}$ can now scatter with numerous other phonons
${\mathbf{q}'\nu'}$ at the same energy 
$\hbar\omega_{{\bf q'}\nu'}\sim0.2$~eV
(the scattering is elastic).
Since such phonons are distributed
in a large zone of the BZ, they are not directly coupled
with conducting electrons and are, thus, mostly not
``hot'' (their occupation is thermalized and very small). In
practice, isotopic disorder allows the scattering of a hot phonon
with other cold phonons adding a deexcitation
channel and, thus, increasing the thermalization rate
from $\tau_{pp}^{-1}$ to $\tau_{pp}^{-1}+\tau_{id}^{-1}$.

\paragraph*{Computation of the isotopic-disorder scattering-times.---}
Let us consider an isotopically enriched tube, with
disorder parameter $x$. $m$
($m+\Delta m$) is the mass if the ${{}^{12}C}$ (${{}^{13}C}$)
isotope. The atomic mass $m_i$ is $m+\Delta m$ with
probability $x$, being in average $\overline{m}=m+x\Delta m$.
In perturbation theory with respect to $\Delta m/m$,
the lifetime ${\tau^{\mathbf{q}\nu}_{id}}$ due
to isotopic disorder of a phonon {\bf q}$\nu$ with
pulsation $\omega_{{\bf q},\nu}$ is given by the Fermi
golden-rule ~\cite{Tamura,Cardona}
\begin{eqnarray}
(\tau^{\mathbf{q}\nu}_{id})^{-1}&=&g_2(x)
\frac{\pi}{2N_q}\omega^2_{\mathbf{q}\nu}
\sum_{\mathbf{q'},\nu'} 
\delta_{\eta}(\omega_{\mathbf{q}\nu}-\omega_{\mathbf{q}',\nu'}) \nonumber \\
& & \times  \sum_{s} |\mathbf{e}^* _{\mathbf{q},\nu}(s) \cdot \mathbf{e}_{\mathbf{q}'\nu'}(s)|^2,
\label{eq1}
\end{eqnarray}
where the sum is performed on $N_q$ ${\bf q}$-points, $s$ is an atomic index,
$\mathbf{e}_{\mathbf{q}\nu}(s)$ is the phonon polarization normalized as
${\sum_s|\mathbf{e}_{\mathbf{q}\nu}(s)|^2 =1 }$ in the unit-cell and
${\delta_{\eta}}$ is a Lorentzian with full-width at half maximum $\eta$, which
is determined self-consistently from 
$ \eta=[\tau^{{\bf q}\nu}_{id}(\eta)]^{-1}$.
${g_2(x)}$ describes  the  mass disorder and is
\begin{eqnarray}
g_2(x) = \frac{\langle(m_i-\overline{m})^2\rangle}{\overline{m}^2}=
x(1-x)\left(\frac{\Delta m}{\overline{m}}\right)^2.
\label{eq2}
\end{eqnarray}
$\omega_{{\bf q},\nu}$ and ${\bf e}_{{\bf q}\nu}$ are
computed from DFT~\cite{DFPT,note03}.

To verify the validity of Eq.~\ref{eq1}, $\tau^{{\bf q}\nu}_{id}$
are also determined exactly, following Ref.~\cite{vast00}.
Let us consider a
super-cell containing $N$ atoms and let us call ${\bf D}={\bf
M}^{-1/2}{\bf K}{\bf M}^{-1/2}$, where {\bf M} is the diagonal $N\times
N$ matrix describing the random distribution of the masses
($M_{ij}=m_i\delta_{ij}$) and {\bf K} are the interatomic harmonic
force constants.  The phonon spectral-function is given by:
\begin{equation}
A(\mathbf{q}\nu,\omega) = 
-\frac{1}{\pi} \mathrm{Im}\,\left\langle V_{\mathbf{q}\nu}
\right|\frac{2 \omega}{(\omega+i\xi)^2-\mathbf{D}}\left| V_{\mathbf{q}\nu} \right\rangle 
\label{eq3}
\end{equation}
where $V_{\mathbf{q}\nu}$ are the atom displacements of the
unperturbed ${\mathbf{q}\nu}$~phonon (normalized to 1 in the super-cell)
and $\xi$ is a small positive number.
Using the DFT force-constants~\cite{DFPT,note03}
and computing $A({\bf q}\nu,\omega)$ with a Lanczos method~\cite{vast00},
$\tau_{id}^{{\bf q}\nu}$ are obtained as the inverse of the full-width at half
maximum of $A({\bf q}\nu,\omega)$~\cite{note04}.

\begin{figure}[h!]
\centerline{\includegraphics[width=80mm]{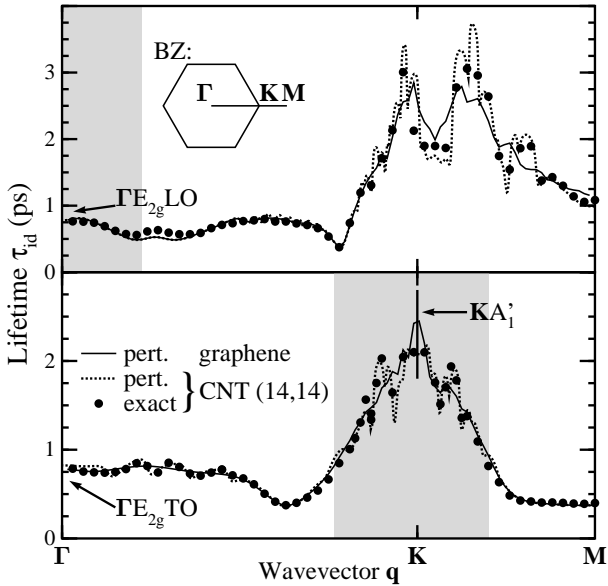}}
\caption{Phonon lifetimes due to isotopic disorder (${x=0.5}$)
in a (14,14) CNT and in graphene computed with the perturbative (pert.)
and exact approach of Eqs.~\ref{eq1} and ~\ref{eq3}, respectively.
The phonons in the two panels correspond to the highest optical branches
of graphene including the ${\bm \Gamma}$E$_{2g}$LO and {\bf K}A$'_1$
modes.
The grey areas correspond to phonons with high occupation (hot phonons)
at 1~Volt.}
\label{fig3}
\end{figure}

$\tau_{id}$ are computed for several
tubes with different diameters $d$ with both Eqs.~\ref{eq1} and ~\ref{eq3}.
Fig.~\ref{fig3} shows the results for the two phonon branches relevant
in electronic transport for a $d=2$~nm tube.
The results from the two methods
are very similar. Thus, the perturbative approach (Eq.~\ref{eq1})
is a good approximation to the exact one (Eq.~\ref{eq3}).
Moreover, the results obtained for a $d=2$~nm nanotube are very similar
to those for two-dimensional graphene. Thus, the results for tubes
with $d>$ 2~nm are not strongly dependent on the diameter.

It is now important to remark that the $\tau_{id}$ for the two hot-phonons
($\tau_{id}^{\bm \Gamma}$ and $\tau_{id}^{\bf K}$ in Fig.~\ref{fig1})
are smaller than the corresponding $\tau_{pp}$. Thus, the new
thermalization channels (due to isotopic disorder) are more efficient
than those present in a defect-free tube (due to anharmonicity)
and are expected to influence the  IV curve.

\paragraph*{Computation of the IV curve.---}
To quantify the impact of isotopic disorder (ID) on the high-bias
electrical resistivity, we use the BTE as in Ref.~\cite{lazzeri06a}.
The ID scattering rate of an initial {\bf q}$\nu$ phonon 
can be decomposed as a sum of scattering rates into the other 
{\bf q}$'\nu'$ phonons, by rewriting Eq.~\ref{eq1} as
$(\tau^{{\bf q}\nu}_{id})^{-1}=\sum_{{\bf q}'\nu'}
\tau_{{\bf q}\nu\rightarrow {\bf q}'\nu'}^{-1}$.
The thermalization process due to ID
is then taken into account by adding to the collision
term for the phonons (Eq.~(4) of Ref.~\cite{lazzeri06a})
${\left[\partial_t n^{\nu}(\mathbf{q})\right]_{iso} =
\sum_{\mathbf{q}'\nu'} \tau_{ \mathbf{q}\nu\rightarrow \mathbf{q}'\nu'
}^{-1} ( n^{\nu'}(\mathbf{q}')- n^{\nu}(\mathbf{q}))}$, where
${n^{\nu}(\mathbf{q})}$ is the occupation of the ${\mathbf{q}\nu}$
phonon. 

\begin{figure}[h!]
\centerline{\includegraphics[width=70mm]{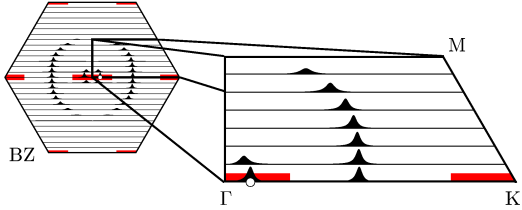}}
\caption{(Color online) Transition rates ${\tau_{\mathbf{q}\nu\rightarrow \mathbf{q}'\nu'}^{-1}}$ 
from an initial ${\mathbf{q}\nu}$~phonon (white dot)
into  ${\mathbf{q}'\nu'}$~phonons (see the text) due to isotopic disorder ($x=0.5$).
Horizontal lines are the wavevector in graphene Brillouin zone (BZ)
compatible with the (14,14) CNT. Filled curves 
are proportional to ${\tau_{\mathbf{q}\nu\rightarrow \mathbf{q}'\nu'}^{-1}}$ 
The red zones correspond to phonons with high occupation (hot phonons)
at 1~Volt.}
\label{fig4}
\end{figure}

First, Fig.~\ref{fig4} reports as example the ID transition rates
for a particular hot-phonon {\bf q}$\nu$.
The majority of the ${\bf q}'\nu'$ states are not ``hot'' (we checked that
more than the 90\% of the scattering rate is due to phonons which are not
``hot''). Second, for each of the ${\bf q}'\nu'$ phonons,
$\tau_{{\bf q}\nu\rightarrow {\bf q}'\nu'}$
is much larger than their anharmonic scattering rate
(as obtained in Ref.~\cite{bonini07}).  
A consequence of these two facts is that
the population of the phonons which are not directly
coupled to conducting electrons, $n^{\nu'}({\bf q'})$,
is negligible and one can rewrite
${\left[\partial_t n^{\nu}(\mathbf{q})\right]_{iso} \sim -
(\sum_{\mathbf{q}'\nu'} \tau_{\mathbf{q}\nu \rightarrow
\mathbf{q}'\nu' }^{-1}) n^{\nu}(\mathbf{q})}$.  That is, one can
directly use Eqs.1-4 of Ref.~\cite{lazzeri06a}, replace the
thermalization time ${\tau_{th}^{-1}}$ by
$({\tau_{pp}^{{\bf q}\nu})^{-1}+(\tau_{id}^{{\bf q}\nu})^{-1}}$ 
corresponding to the two hot-phonons and consider the {\bf q} dependence
of $\tau_{id}$~\cite{note_ref02}.

In Fig.~\ref{fig2} we report the IV curve for the isotopically
enriched ($x$=0.5) carbon nanotube.  Indeed, the conduction properties
are ameliorated with respect to the isotopically pure case ($x=0$).
To be quantitative, we notice that
at high-bias the differential resistance $dV/dI$ is almost independent
from the voltage and is linear with respect to the tube length
(lower panel of Fig.~\ref{fig2}). This allows to define a resistivity
$\rho$ and an effective scattering length $l_{sc}$, by
\begin{equation}
\rho=\frac{1}{L}\frac{dV}{dI} = \frac{1}{l_{sc}} \frac{h}{4e^2},
\label{eq4}
\end{equation}
where $h/(2e^2)=12.9$ k$\Omega$ is the quantum of resistance.
From Fig.~\ref{fig2}, the introduction of
isotopic disorder enhances the scattering length ( and thus reduces the
high-bias differential resistance) by a factor three.

\paragraph*{Conclusions.---}
The introduction of isotopic disorder in metallic carbon
nanotubes creates an effective
channel for the thermalization of the hot phonons without introducing
an additional scattering process for the conducting electrons.
This is expected to reduce the actual electronic differential-resistance.
In particular, according to a Boltzmann transport treatment 
(in which all the relevant parameters are obtained from
ab-initio and no semi-empirical parameters
are present which could be tuned to reproduce the experimental results),
isotopic disorder reduces the
high-bias electronic differential-resistance by a factor of three
in tubes with a diameter of 2~nm.
This amelioration of the transport properties is expected to have important
technological consequences in view of the use of metallic carbon nanotubes
as interconnects in tomorrow electronics.
Further improvement could be reached by adding other sources of phonon
scattering wich are transparent to electrons as, e.g., by cycloaddition
functionalization~\cite{lee06}.

After submission of the present work, another paper appeared proposing
isotopic enrichment to reduce hot-phonon population
in GaN transistors~\cite{khurgin08}.
We thank G. Stolz, N. Bonini and N. Marzari for discussions
and aknowledge IDRIS project 081202.

\begin{thebibliography}{99}

\bibitem{yao00} Z. Yao, C. L. Kane, and C. Dekker, Phys. Rev. Lett. {\bf 84}, 2941 (2000).

\bibitem{javey04}
A. Javey {\it et al.}, 
Phys. Rev. Lett. {\bf 92}, 106804 (2004).

\bibitem{park04}
J.-Y. Park {\it et al.}, 
Nano Lett. {\bf 4}, 517 (2004).

\bibitem{lazzeri05}
M. Lazzeri {\it et al.}, 
Phys. Rev. Lett. {\bf 95}, 236802 (2005).  

\bibitem{lazzeri06a}
M. Lazzeri and F. Mauri, Phys. Rev. B {\bf 73}, 165419 (2006). 

\bibitem{oron-carl08}
M. Oron-Carl and R. Krupke, Phys. Rev. Lett. {\bf 100}, 127401 (2008).

\bibitem{bushmaker07}
A.W. Bushmaker, V.V. Deshpande, M.W. Bockrath, and S.B. Cronin,
Nano Lett. {\bf 7}, 3618 (2007).

\bibitem{pop05}
E. Pop {\it et al.}, 
Phys. Rev. Lett. {\bf 95}, 155505 (2005).

\bibitem{note-1}
Within the Born-Oppenheimer approximation, the electronic Hamiltonian
depends on the nuclear charges and positions but not explicitely on the nuclear masses.
If the nuclear motion is harmonic, the average nuclear-positions do not
depend on the nuclear masses.
Thus, isotopic substitution scatters electrons only by mechanisms (which
in general are negligible) beyond the
Born-Oppenheimer approximation and/or beyond the harmonic
approximation for interatomic force-constants and for electron-phonon couplings.
These mechanisms have been studied in literature [see e.g. Sect. 4.2 of
A.P. Zhernov, and A.V. Inyushkin, Physics-Uspekhi {\bf 45} 527-552 (2002)]
with predicted electron scattering-lengths of the order of the millimeter.
These scattering lengths are undetectable with respect to those
due to other sources of electron scattering found in bulk metals and in carbon nanotubes,
where the relevant lenghts range from 10~nm (electron-phonon scattering)
to 1600~nm (elastic scattering)~\cite{park04}.

\bibitem{note00}
In metallic CNTs, the electronic bands cross the
Fermi energy at two equivalent points of the reciprocal
space corresponding to the symmetry points {\bf K} and {\bf K'}
of the graphene BZ~\cite{lazzeri05}.
Because of momentum conservation, the conducting electrons
of the tube can scatter only with phonons with momentum near to
the zone-center ${\bm \Gamma}$ or to {\bf K}.
The phonons mostly involved in this process correspond to the 
E$_{2g}$ at ${\bm \Gamma}$ (the component longitudinal to the
tube axis, LO) and to the A$'_1$ at {\bf K} of graphene, since
these are the phonons with the strongest electron-phonon coupling
~\cite{lazzeri05}.
Their momentum, {\bf q} and {\bf K+q}, is localized in the BZ
region defined by $|{\bf q}|<2Ve/(\hbar v_F)$, where $V$ is the
applied voltage and $v_F$ is the Fermi velocity.

\bibitem{note01}
$\tau_{ep}$ is the average time a conducting electron can travel before
emitting an optical phonon by back-scattering due to electron-phonon
coupling.
It is determined by DFT calculations
~\cite{DFPT,lazzeri05} and agrees very well with Raman measurements
of the phonon $G$-linewidth in graphite, graphene and metallic
CNTs~\cite{lazzeri06b,bonini07}.
$\tau_{pp}$ is determined by the anharmonic phonon-phonon interaction
via DFT in graphene~\cite{bonini07}
and agrees very well with measurements of the phonon decay time
from from ultra-fast spectroscopy measurements~\cite{kampfrath05}.

\bibitem{DFPT}
S. Baroni, S. de Gironcoli, and A. Dal Corso,
Rev. Mod. Phys. B {\bf 73} , 515 (2001).

\bibitem{lazzeri06b}
M. Lazzeri {\it et al.}, 
Phys. Rev. B {\bf 73}, 155426 (2006).

\bibitem{bonini07}
N. Bonini, M. Lazzeri, N. Marzari, and F. Mauri,
Phys. Rev. Lett. {\bf 99}, 176802 (2007).

\bibitem{kampfrath05}
T. Kampfrath {\it et al.},
Phys. Rev. Lett. {\bf 95}, 187403 (2005).

\bibitem{note_ref01}
For isolated nanotubes suspended in vacuum (as in Ref.~\onlinecite{pop05})
the acoustic phonons cannot be considered thermalized with the environment
and experimentally a negative differential resistence can be observed
(see discussion in Ref.~\onlinecite{lazzeri06a}).
This situation is not treated here.

\bibitem{note02}
Ref.~\cite{lazzeri06a} describes the coupled BTE
for the populations of both conducting electrons and
optical phonons. It does
not rely on any further approximation than those inherent to the
Boltzmann one. The optical-phonons evolution is determined by the
electron-phonon scattering times ($\tau_{ep}$) and
the thermalization-time.
Contrary to
Ref.~\cite{lazzeri06a}, here we consider as thermalization times
the anharmonic scattering times ($\tau_{pp}$)
obtained from DFT calculations (see the text and note~\cite{note01}).

\bibitem{BN}
C. W. Chang {\it et al.},
Phys. Rev. Lett. {\bf 97}, 085901 (2006).

\bibitem{Kuzmany1}
V. Z\'olyomi {\it et al.},
Phys. Rev. B {\bf 75}, 195419 (2007).

\bibitem{Kuzmany2}
F. Simon {\it et al.},
Phys. Rev. Lett. {\bf 95}, 017401 (2005).

\bibitem{Cardona}
F. Widulle, J. Serrano, and M. Cardona, Phys. Rev. B {\bf 65}, 075206 (2002). 

\bibitem{Tamura}
S. I. Tamura, Phys. Rev. B {\bf 27}, 858 (1983).

\bibitem{note03}
Force constants are computed in a ${9\times9}$ graphene super-cell
with the same computational details as in~\cite{bonini07} and then refolded
to obtain those of the nanotube.

\bibitem{note04}
The disordered system is simulated with super-cells containing
over 10000~atoms.
Spectral functions are averaged over five different mass distributions and
then fitted with a Lorentzian distribution to obtain the life-time.

\bibitem{vast00}
N. Vast, and S. Baroni, Phys. Rev. B {\bf 61}, 9387 (2000).

\bibitem{note_ref02}
The isotope substitution slightly modifies also
$\omega$, $\tau_{ep}$, and $\tau_{pp}$, which scale as
$\overline{m}^{-1/2}$, $\overline{m}^{1/2}$, and $\overline{m}$, respectively.
We checked that such modifications do not affect the IV curves of Fig.~\ref{fig2}
in a detectable way.

\bibitem{lee06}
Y.S. Lee and N. Marzari, Phys. Rev. Lett. {\bf 97}, 116801 (2006).

\bibitem{khurgin08}
J.B. Khurgin, D. Jena, and Y.J. Ding,
Appl. Phys. Lett. {\bf 93}, 032110 (2008).

\end{thebibliography}
\end{document}